\documentclass[12pt]{iopart}
\usepackage{iopams}
\usepackage{setstack}
\usepackage{graphicx}

\begin{document}

\title[Quantum tunneling as a classical anomaly] {Quantum tunneling as a
classical anomaly}

\author[Bender and Hook]{Carl~M~Bender${}^\ast$ and
Daniel~W~Hook${}^{\ast\dag}$}

\address{${}^\ast$Department of Physics, Washington University, St. Louis, MO
63130, USA \\{\footnotesize{\tt email: cmb@wustl.edu}}}

\address{${}^\dag$Theoretical Physics, Imperial College, London SW7 2AZ,
UK\\ {\footnotesize{\tt email: d.hook@imperial.ac.uk}}}

\date{\today}

\begin{abstract}
Classical mechanics is a singular theory in that real-energy classical particles
can never enter classically forbidden regions. However, if one regulates
classical mechanics by allowing the energy $E$ of a particle to be complex, the
particle exhibits quantum-like behavior: Complex-energy classical particles can
travel between classically allowed regions separated by potential barriers. When
${\rm Im}\,E\to0$, the classical tunneling probabilities persist. Hence, one can
interpret quantum tunneling as an anomaly. A numerical comparison of complex
classical tunneling probabilities with quantum tunneling probabilities leads to
the conjecture that as ${\rm Re}\,E$ increases, complex classical tunneling
probabilities approach the corresponding quantum probabilities. Thus, this work
attempts to generalize the Bohr correspondence principle from classically
allowed to classically forbidden regions.
\end{abstract}

\submitto{\JPA}

Classical mechanics is a singular limit of quantum mechanics. (The limit
$\epsilon\to0$ is {\it singular} if an abrupt change occurs at $\epsilon=0$.
For example, $\lim_{\epsilon\to0}\epsilon x^5+x=1$ is singular because four of
its five roots abruptly disappear at $\epsilon=0$.) The classical limit $\hbar\to0$
of the time-independent Schr\"odinger equation $\hbar^2\psi''(x)=[V(x)-E]\psi(x
)$ is singular because at $\hbar=0$ it is no longer possible to impose initial
or boundary conditions on the wave function $\psi(x)$. Moreover, while quantum
particles are able to enter classically forbidden regions, these particles
abruptly lose this ability at $\hbar=0$. Thus, the phenomenon of tunneling seems
to be entirely quantum mechanical, and one may not ask such classical questions
as, Which path does the particle follow while tunneling?

It may be possible to recover some features of a theory that were abruptly lost
in a singular limit, like the ability of a particle to enter a classically
forbidden region. To do so, one introduces a {\it regulator}, which is then
removed in a careful limiting process. The features that remain after the
regulator is removed are referred to as {\it anomalies}. (The axial anomaly in
quantum field theory can be obtained by using dimensional regulation.)

Particle trajectories in conventional classical mechanics are real functions of
time, but recent studies of the complex solutions to the classical equations of
motion (Hamilton's equations) have shown that real-energy classical particles
may leave the real axis and travel through complexified coordinate space
\cite{R1,R2,R3,R4,R5,R6,R7,R8,R9,R10,R11,R12}. While these complex classical
trajectories may pass through classically forbidden regions on the real axis,
complex classical mechanics is still a singular theory because there is no
tunneling; that is, no complex path runs from one classically allowed region on 
the real axis to another.

Complex classical mechanics may be regulated by taking the energy of a particle
to be complex. Surprisingly, a complex-energy classical particle exhibits
qualitative features normally associated with a quantum particle: Such a
classical particle can exhibit tunneling-like behavior in which it travels from
one classically allowed region to another classically allowed region even though
these two regions are separated on the real axis by a classically forbidden
region \cite{R20,R22}. Furthermore, a complex-energy classical particle in a
periodic potential exhibits quantum-like behavior; there are sharply defined
energy bands separated by gaps. In these energy bands the classical particle
exhibits a kind of resonant tunneling \cite{R20}. The time-energy uncertainty
principle supports our choice of regulator. This uncertainty principle implies
that a precise measurement of the energy of a particle in a finite time interval
is impossible; some uncertainty $\Delta E$ is associated with such a
measurement. In this paper we allow $\Delta E$ to be complex.

This paper explores the connection between standard quantum mechanics and
complex-energy classical mechanics at a {\it quantitative} level. We show that
as the regulator is removed (${\rm Im}\,E\to0$), the tunneling behavior of
complex-energy classical mechanics {\it persists}. Thus, we recover tunneling as
a kind of anomaly.

To compare the behavior of quantum particles and complex classical particles,
we use quartic and sextic asymmetric double-well potentials and compute
numerically the relative probabilities of finding these particles in each well.
We find that as the number of nodes in the quantum-mechanical eigenfunction
increases, the quantum and classical probabilities approach one another. Thus,
we demonstrate that the Bohr correspondence principle, which has recently been
generalized to the complex domain \cite{R23,R25}, actually applies to tunneling
phenomena.

\begin{figure}
\begin{center}
\includegraphics[scale=0.34, bb=0 0 1000 1179]{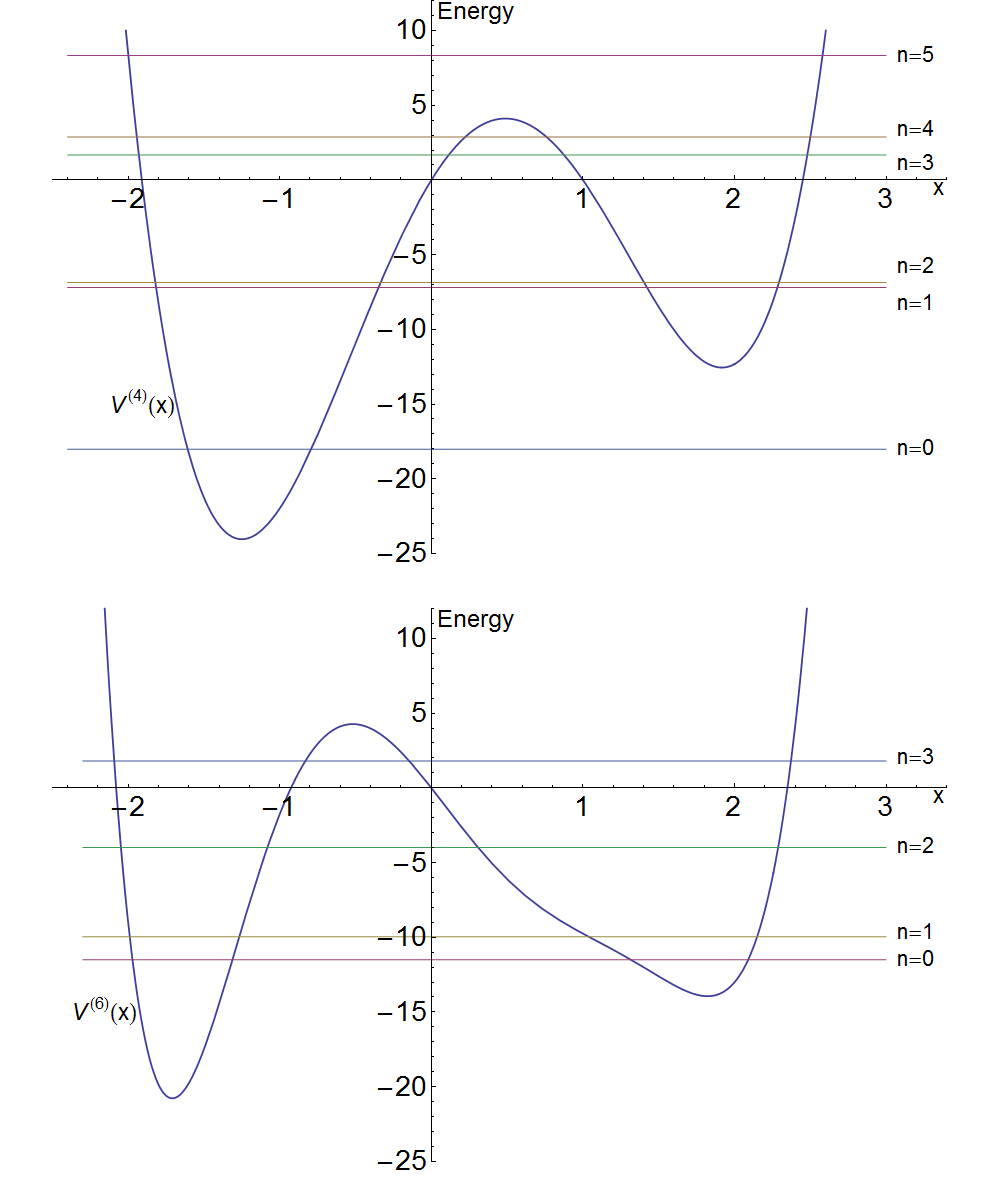}
\end{center}
\caption{Upper panel: Quartic asymmetric double-well potential $V^{(4)}(x)$ in
(\ref{e1}) and the first six quantum energy levels. There are minima at $x=-
1.2499$, where $V^{(4)}=-24.0384$, and at $x=1.9165$, where $V^{(4)} =-12.5501$.
The peak of the barrier is at $x=0.4884$, where $V^{(4)}=4.1144$. The
ground-state energy $E_0$ lies below the bottom of the right potential well.
The next four energy levels lie between the bottom of the right potential well
and the top of the barrier. The sixth energy level $E_5$ lies above the barrier.
Lower panel: Sextic asymmetric double-well potential $V^{(6)}(x)$ in (\ref{e2}).
The minima of the wells are at $x=-1.7083$, where $V^{(6)}=-20.7710$, and at $x=
1.8215$, where $V^{(6)}=-13.9373$. The peak of the barrier is located at $x=-
0.5184$, where $V^{(6)}=4.2731$.}
\label{F1}
\end{figure}

The quartic double-well potential
\begin{equation}
V^{(4)}(x)=\textstyle{\frac{7}{2}}x(x-1)\left(x+\textstyle{\frac{191}{100}}
\right)\left(x-\textstyle{\frac{49}{20}}\right)
\label{e1}
\end{equation}
is shown in the upper panel in Fig.~\ref{F1}. The first six quantum energy
levels of $H=p^2+V^{(4)}(x)$, where we have set $\hbar=1$, are $E_0=-18.0182$
(below the bottom of the right well), $E_1=-7.1879$, $E_2=-6.8595$, $E_3=
1.6806$, and $E_4=2.8845$ (above the bottom of the right well and below the peak
of the barrier, and $E_5=8.3312$ (above the barrier).

We plot the sextic potential
\begin{equation}
V^{(6)}(x)=x^6-2x^5-4x^4+11x^3-\textstyle{\frac{11}{4}}x^2-13x
\label{e2}
\end{equation}
in the lower panel in Fig.~\ref{F1}. The exact ground-state energy for $H=p^2+
V^{(6)}(x)$ with $\hbar=1$ is $E_0=-23/2$. (This exact value of $E_0$ provides a
benchmark confirming that our numerical calculations are accurate to better than
13 decimal places.) The first three excited quantum states have energies $E_1=-
9.9690$, $E_2=-3.9819$, and $E_3=1.8095$.

\begin{figure}[t!]
\begin{center}
\includegraphics[scale=0.34, bb=0 0 1000 1265]{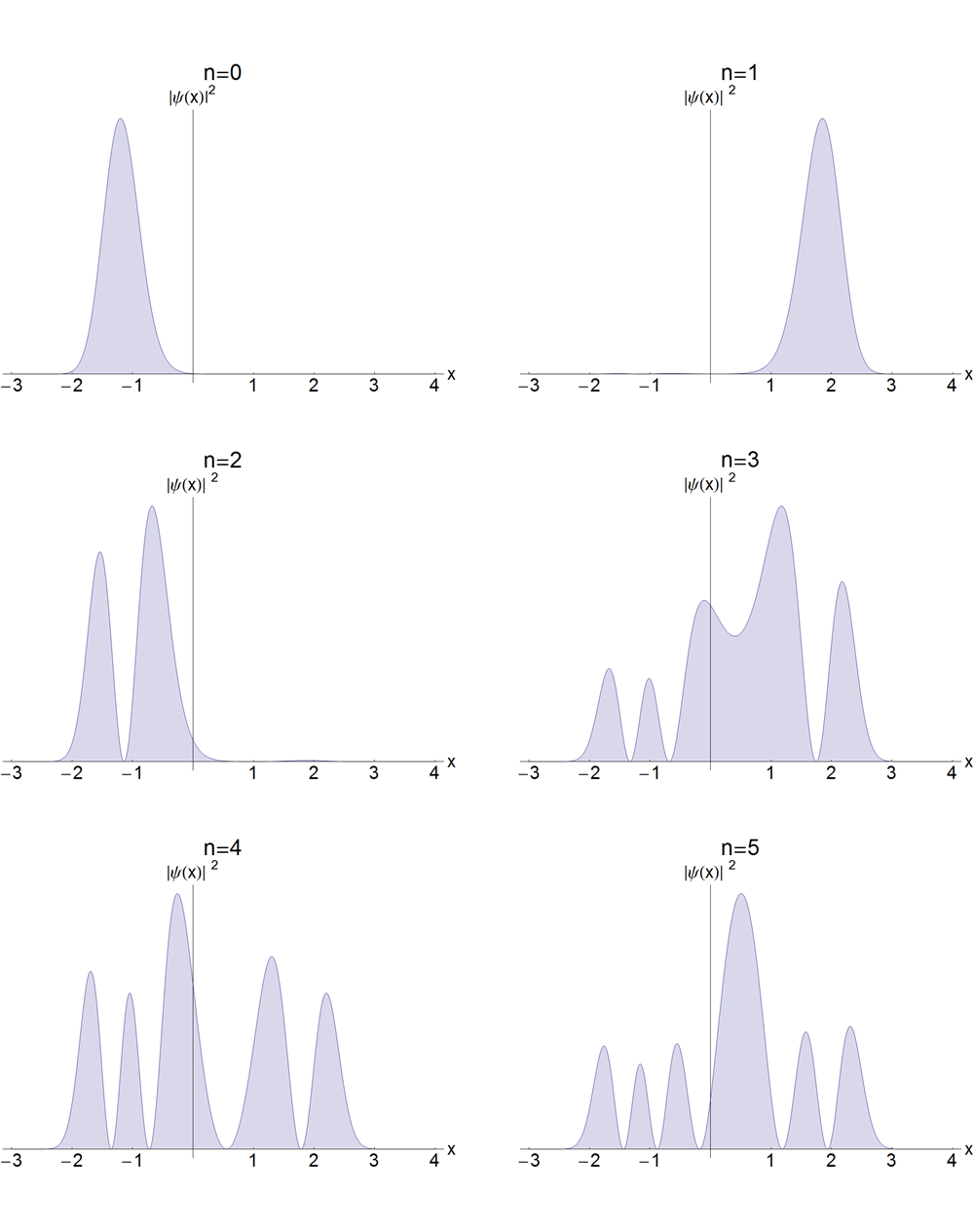}
\end{center}
\caption{Quantum probability density for a particle in the $n$th eigenstate in
the quartic potential $V^{(4)}(x)$ in (\ref{e1}). The probability $P_{{\rm right
},n}^{\rm quant}$ in (\ref{e3}) of finding a particle in the $n$th eigenstate to
the right of the barrier is determined by integrating the $n$th probability
density.}
\label{F2}
\end{figure}

The eigenfunction $\psi_n(x)$ corresponding to the energy $E_n$ has $n$ nodes.
In Fig.~\ref{F2} we plot the probability density $|\psi_n(x)|^2$ for a particle
in $V^{(4)}(x)$ for $n=0,\,1,\,\ldots,\,5$. We integrate $|\psi_n(x)|^2$ to
determine the probability $P_{{\rm right},n}^{\rm quant}$ of finding the
particle to the right of the barrier \cite{R30}:
\begin{eqnarray}
P_{{\rm right},1}^{\rm quant}=99.5933\%,&\qquad&P_{{\rm right},2}^{\rm quant}=
0.4316\%,\nonumber\\
P_{{\rm right},3}^{\rm quant}=59.7584\%,&\qquad& P_{{\rm right},4}^{\rm quant}=
40.7689\%.
\label{e3}
\end{eqnarray}
For the sextic potential $V^{(6)}(x)$ in (\ref{e2}) we find that for the first
four quantum states the probabilities of finding the particle to the right of
the top of the barrier are
\begin{eqnarray}
P_{{\rm right},0}^{\rm quant}=0.0391\%,&\qquad& P_{{\rm right},1}^{\rm quant}=
99.9986\%,\nonumber\\
P_{{\rm right},2}^{\rm quant}=99.8651\%,&\qquad& P_{{\rm right},3}^{\rm quant}=
78.7223\%.
\label{e4}
\end{eqnarray}

\begin{figure}
\begin{center}
\includegraphics[scale=0.34, bb=0 0 1000 954]{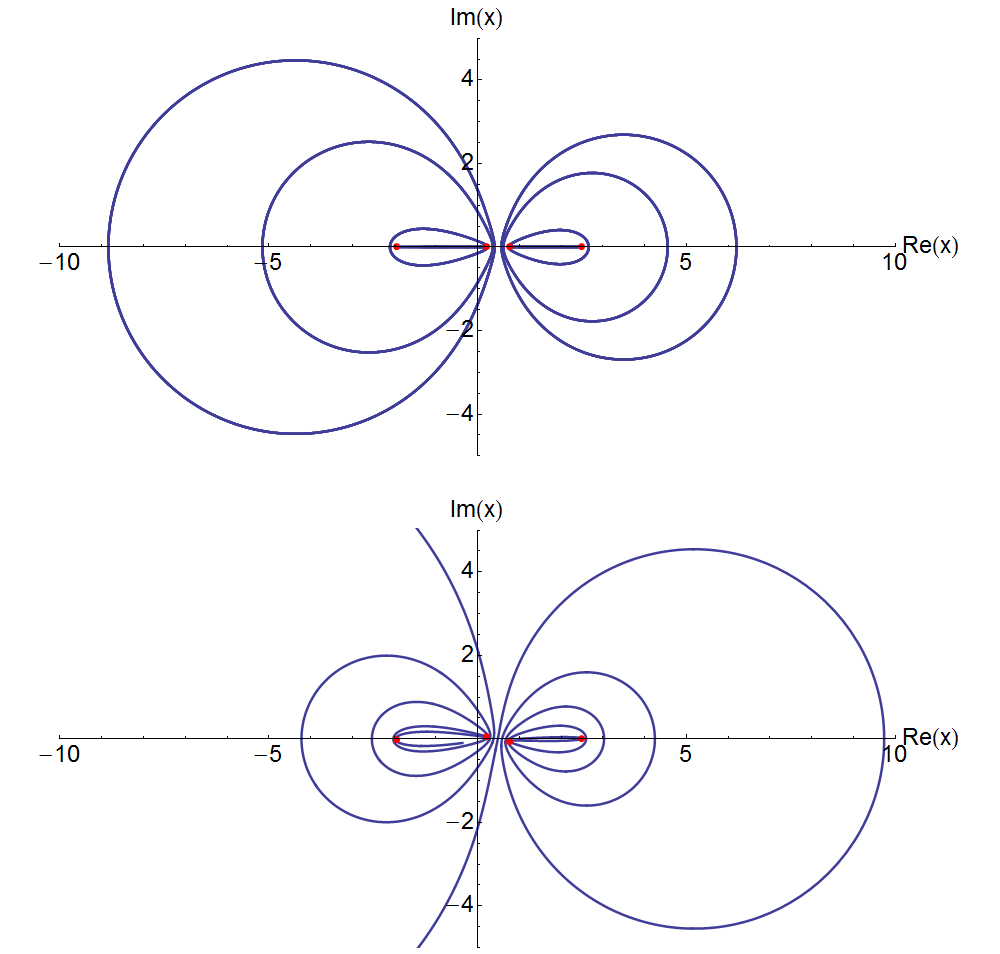}
\end{center}
\caption{Upper panel: Periodic trajectories in the complex-$x$ plane of a
classical particle of real energy $E_4=2.8845$ in the quartic potential
(\ref{e1}). The dots indicate the four turning points. There is no
tunneling-like behavior; no trajectory runs from one classically
allowed region to the other. Lower panel: Classical trajectory in the
complex-$x$ plane of a particle of complex energy $E=2.8845+0.5i$ in the same
potential (\ref{e1}). Unlike the closed trajectories in the upper panel, this
classical trajectory is open. The particle begins at $x=1$ and traces an outward
anticlockwise spiral around the right pair of turning points. It then crosses
to the left of the peak of the barrier and spirals inward and clockwise around
the left pair of turning points. Next, it crosses the real axis between the left
turning points and spirals outward and anticlockwise around the left pair of
turning points. The time for the displayed trajectory is $t=8$. For $t>8$, the
particle continues to spiral inward and outward as it oscillates from well to
well. The particle exhibits this deterministic tunneling-like behavior for all
time, but the trajectory never crosses itself. This trajectory behaves as if it
were controlled by a pair of strange attractors.}
\label{F3}
\end{figure}

We now describe the behavior of a classical particle in the quartic potential
(\ref{e1}) for the case in which $x$ is complex. Such a particle may have either
real or complex energy, and we begin by considering the complex trajectories of
a real-energy classical particle. Let us take the energy of a classical particle
in $V^{(4)}$ to be $E_4=2.8845$. There are two real turning points at $x=
-1.9428$ and $0.2234$, which bound the classically allowed region to the left of
the barrier and two turning points at $x=0.7596$ and $2.4998$, which bound the
classically allowed region to the right of the barrier. If the initial position
of the particle is real and in the classically allowed region in the right (or
left) well, the trajectory of the particle oscillates on the real axis between
the right (or left) pair of turning points. However, if the particle is
initially in a classically forbidden region on the real axis, the particle
leaves the real axis and travels in a closed periodic orbit in the complex-$x$
plane [see Fig.~\ref{F3} (upper panel)]. The orbits in the complex-$x$ plane
enclose the classically allowed regions on the real axis and never cross the
vertical line ${\rm Re}\,x=0.4884$, which passes through the peak of the
barrier. The periods $T$ of all closed orbits, both on the left and right side
of Fig.~\ref{F3}, are the same and are given by
$$T=\int_{x_{\rm left}}^{x_{\rm right}}\frac{dx}{\sqrt{E-V^{(4)}(x)}},$$
where $x_{\rm left}$ and $x_{\rm right}$ are the left and right turning points
in either well and $E$ is the (real) classical energy of the particle. For
the case in which the energy $E=E_4=2.8845$, the period $T=0.8464$. The particle 
in the upper panel of Fig.~\ref{F3} cannot travel from one classically allowed 
region to the other classically allowed region, so there is no tunneling effect 
when the classical energy is exactly real.

\begin{figure}
\begin{center}
\includegraphics[scale=0.34, bb=0 0 1000 973]{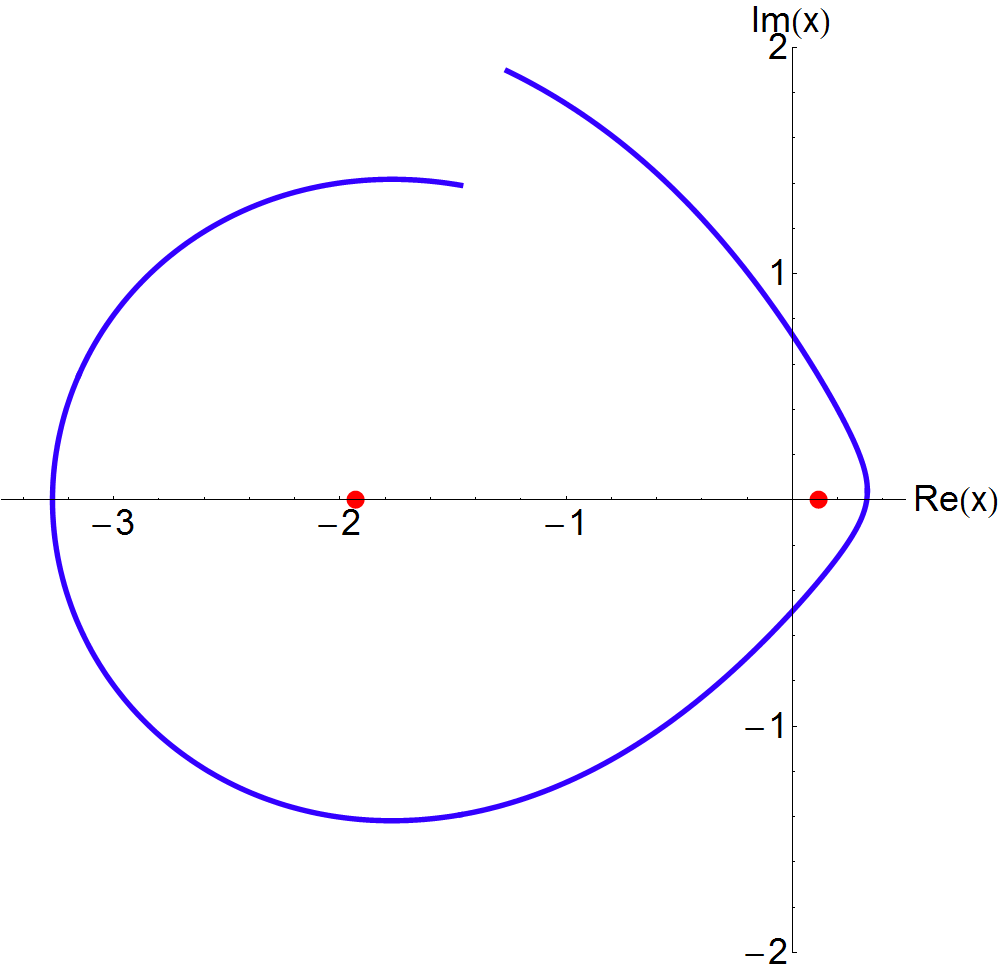}
\end{center}
\caption{Complex path of a particle of complex energy $E=2.8845+0.25i$. The time
of this path is taken to be $t=T=0.8464$, which is the same as the period of a
classical particle of real energy $E=2.8845$. Note that the particle in this
figure makes an approximately full angular revolution around the two two turning
points, which are denoted by dots. However, the path is not closed.}
\label{F4}
\end{figure}

We now explain how deterministic classical systems can produce results that are
numerically comparable to the quantum results in (\ref{e3}) and (\ref{e4}). If
the energy $E$ of the classical particle is complex, the classical trajectory of
such a particle is not in general closed \cite{R20}. In Fig.~\ref{F4} we plot
the complex path of a classical particle of energy $E=2.8845+0.25i$ for the
time interval from $t=0$ to $t=T=0.8464$. Observe that while the particle
executes a loop of nearly $360^\circ$ around the two turning points, the
trajectory of the particle is no longer closed. If this particle is initially in
a classically allowed region, it spirals outward around the pair of turning
points that bound the region. However, the particle does not drift off to
infinity. Rather, it crosses the vertical line that passes through the top of
the barrier and spirals into the other well. This behavior is shown in
Fig.~\ref{F3} (lower panel), where a particle of energy $E=2.8845+0.5i$
alternately visits both potential wells. This classical particle executes
deterministic {\it tunneling} from well to well.

\begin{figure}
\begin{center}
\includegraphics[scale=0.80, bb=0 0 358 399]{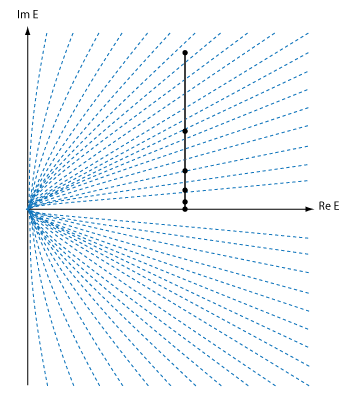}
\end{center}
\caption{Schematic drawing of the special complex energies for which the
particle motion in the quartic potential (\ref{e1}) is periodic for any choice
of initial condition. These energies lie on asymptotically straight dashed
curves that approach the origin in the complex-$E$ plane. The energies form a
set of measure zero; there are an infinite number of such curves and these
curves are dense in the complex-$E$ plane. The regulation procedure used in this
paper establishes a connection between complex classical mechanics and
quantum mechanics by taking a sequence of complex classical energies in
which the imaginary part tends to 0. This sequence is indicated by dots on the 
solid vertical line. Although this line intersects the dashed lines infinitely 
many times, the numerical procedure remains unaffected.}
\label{F5}
\end{figure}

There is an unexpected subtlety in the behavior of complex-energy classical
particles that should not be overlooked. While the spiral motion shown in the
lower panel in Fig.~\ref{F3} is typical of complex-energy classical particles,
it has recently been discovered that for a dense set of measure zero of complex
classical energies the classical paths are periodic \cite{R22}. The complex
energies for which the classical motion remains closed and periodic lie on an
infinite number of nearly straight lines that eminate from the origin in the
complex-energy plane. A schematic representation of these lines is shown in
Fig.~\ref{F5}.

The key advance in this paper is the numerical observation that the
probabilities associated with the tunneling of a classical particle having
complex energy {\it persist as the regulator ${\rm Im}\,\Delta E$ tends to zero}
even though there is no classical tunneling when ${\rm Im}\,E=0$. The upper
panel in Fig.~\ref{F3} shows that a classical particle having real energy does
not exhibit tunneling behavior. Yet, a classical particle having ${\rm Im}\,E
\neq0$ typically {\it does} exhibit tunneling behavior, and the corresponding
probabilities can be compared in the limit as ${\rm Im}\,E\to0$ with the quantum
tunneling probabilities in (\ref{e3}) and (\ref{e4}).

To demonstrate the persistence of classical tunneling in the limit ${\rm Im}\,E
\to0$, we compute the classical trajectory for long times \cite{ZZ} and
determine the
fraction of time that the classical particle spends to the left and to the
right of the line ${\rm Re}\,x=0.4884$ (the location of the peak of the
potential barrier). As ${\rm Im}\,E$ gets smaller, the fraction of time spent to
the right (and to the left) of the barrier approaches a constant.\footnote{
Figure \ref{F4} shows that each angular revolution in the complex-energy spiral
takes approximately the same time regardless of whether the spiral encircles the
left or the right pair of turning points. The left-right asymmetry in the well
gives rise to different left and right winding numbers in the spiral path.}
Take the real part of the classical energy to be the fourth excited quantum
energy ${\rm Re}\,E=E_4=2.8845$ and take ${\rm Im}\,E=2^{-k}$. For $k=0,\,1,\,2,
\,3,\,4,\,5$ the chance of finding the classical particle to the right of the
barrier is $50.4\%$, $53.9\%$, $55.0\%$, $53.1\%$, $52.8\%$, $53.3\%$. This
sequence approaches a limiting anomalous value of about $53\%$, which exceeds
$50\%$ in accordance with the heuristic semiclassical argument in
Ref.~\cite{R30}. A classical version of the time-energy uncertainty principle
applies here: As the imaginary part of the energy gets smaller, it takes more
time for the particle to oscillate between the wells \cite{R20}, and thus this
extrapolation procedure for determining the classical tunneling probabilities
requires more computer time.

Classical tunneling persists in the limit as ${\rm Im}\,E\to0$ even though there
is no classical tunneling at ${\rm Im}\,E=0$; Fourier series exhibit strongly
analogous behavior. Take the function $f(x)=1$ on $[0,\pi]$. Although $f(0)\neq
0$ and $f(\pi)\neq 0$, we can represent $f(x)$ as the Fourier sine series
$\frac{4}{\pi}\sum_{k=0}^\infty\frac{1}{2k+1}\sin[(2k+1)x]$. The partial sum
$S_K(x)=\sum_{k=0}^K a_k\sin(kx)$ converges to $f(x)$ as $K\to\infty$ on the
open interval $(0,\pi)$ because $f(x)$ is continuous. All terms in $S_K(x)$ {\it
vanish} at 0 and $\pi$, but we can still recover the nonzero value of $f(0)$ and
$f(\pi)$ from $S_K(x)$. The extrapolation procedure used above for classical
tunneling probability can be applied to Fourier series to determine $f(0)$. We
take twice as many terms in $S_K(x)$ as $x$ is halved and thereby circumvent the
problem of nonuniform convergence (the Gibbs phenomenon): We evaluate $S_K(x)$
at $K=100\times 2^k$ and $x=2^{-k}$ for $k=0,\,\ldots,5$. The numerical values
of $S_K(x)$ are $S_{100}=0.997\,776,\,S_{200}=0.996\,704,\,S_{400}=0.997\,293,\,
S_{800}=0.997\,818,\,S_{1600}=0.998\,128,\,S_{3200}=0.998\,292$. We infer that
$f(0)=1$.

In analogy with the extrapolation scheme used for Fourier series, we determine
the classical probability of finding the particle to the right of the peak of
the barrier for a sequence of energies in which the real part of the energy is
held fixed and the imaginary part of the energy tends to zero. For a small but
fixed imaginary classical energy, the classical probability approaches the
quantum probability as the real part of the energy increases. For example, when
${\rm Im}\,E=1/4$ and ${\rm Re}\,E=E_n$, where $E_n$ is the $n$th quantum
eigenenergy for the quartic potential $V^{(4)}(x)$ in (\ref{e1}), we obtain the
following classical probabilities for finding the classical particle in the
right well:
\begin{eqnarray}
P_{{\rm right},1}^{\rm class}=55.4\%,&\qquad&P_{{\rm right},2}^{\rm class}=55.0
\%,\nonumber\\
P_{{\rm right},3}^{\rm class}=54.3\%,&\qquad& P_{{\rm right},4}^{\rm class}=55.0
\%.
\label{e6}
\end{eqnarray}
The quantum probabilities $P_{{\rm right},n}^{\rm quant}$ in (\ref{e3})
oscillate about these classical probabilities and are in good agreement when
$n=3$ and $n=4$. For deeper double-well potentials, the classical probabilities
continue to approach the quantum probabilities as $n$ increases.

\begin{figure}
\begin{center}
\includegraphics[scale=0.34, bb=0 0 1000 1272]{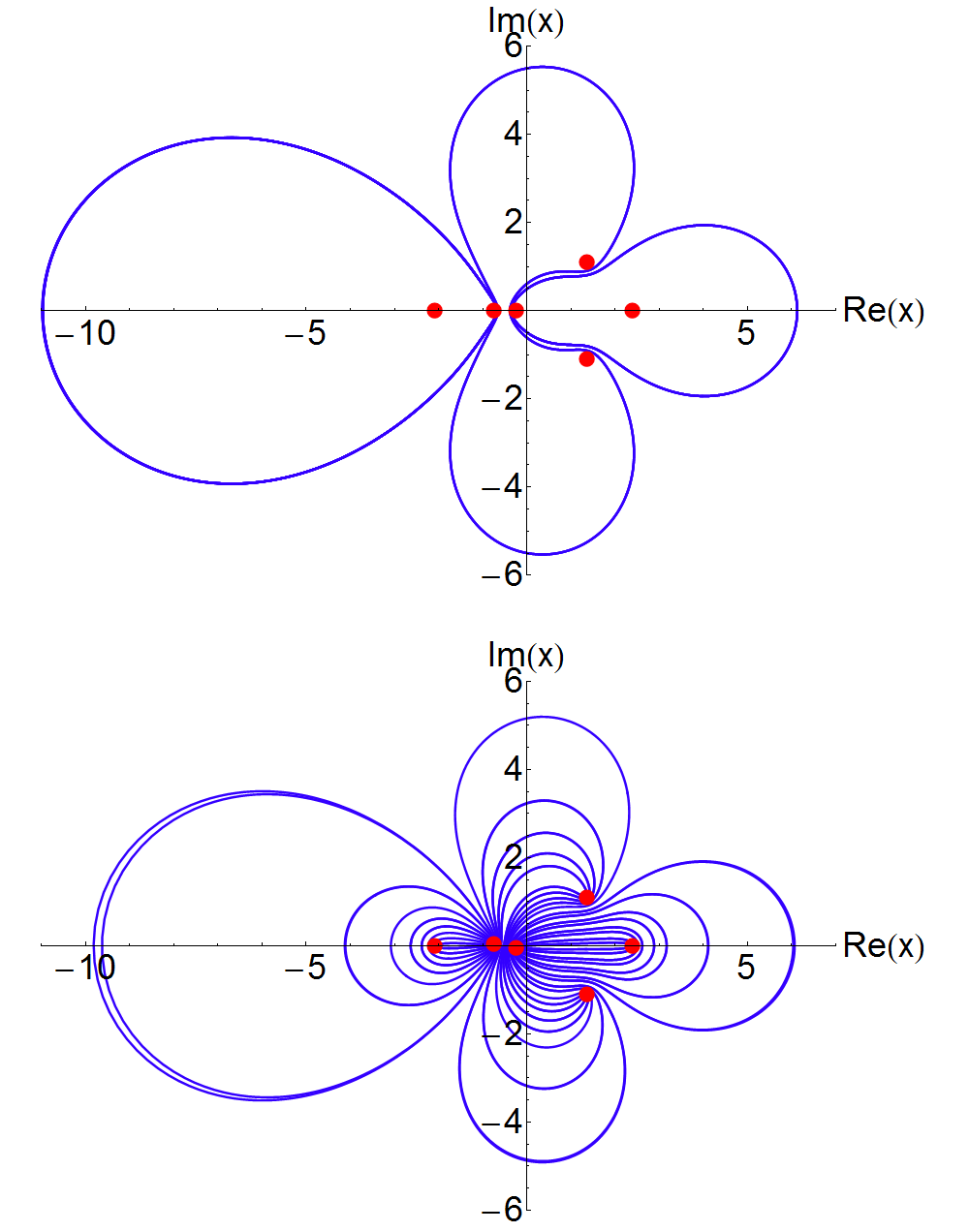}
\end{center}
\caption{Complex classical trajectories for a particle subject to the sextic
potential (\ref{e2}). The energy of the particle in the upper panel is real and
three closed nonintersecting trajectories for this particle are shown. The
energy of the particle in the lower panel is complex and a single open
trajectory is plotted.}
\label{F6}
\end{figure}

For the sextic potential $V^{(6)}(x)$ in (\ref{e2}) there are six turning
points. When the classical energy is real and between the minimum of the lower
well and the top of the barrier, the turning points group into three pairs, one
pair on the real axis to the left of the barrier, a second pair on the real axis
to the right of the barrier, and a complex-conjugate pair that is associated
with the complex extension of the barrier. If we take the classical energy to be
real, say $E_n$, we cannot observe tunneling because the classical orbits are
closed and periodic and just encircle the pairs of turning points, as shown
in the upper panel of Fig.~\ref{F6}.

When the classical energy $E$ is complex, the classical path is no longer
closed. A particle trajectory beginning at $x=1$ in the right well spirals
outward and eventually circles around the barrier turning points. The trajectory
does not always penetrate to the left well; sometimes the particle is ejected
and falls back into the right well. However, after sufficiently many tunneling
attempts the particle spirals inward around the left pair of turning points, as
shown in the lower panel of Fig.~\ref{F6}. Taking ${\rm Im}\,E=1/16$ and ${\rm
Re}\,E=E_n$ ($n=0,\,1,\,2,\,3$), we find that as the real part of the energy
increases, the classical tunneling probabilities $P_{{\rm right},n}^{\rm class}$
listed below approach the quantum probabilities $P_{{\rm right},n}^{\rm quant}$
in (\ref{e4}):
\begin{eqnarray}
P_{{\rm right},0}^{\rm class}=91.7\%,&\qquad&P_{{\rm right},1}^{\rm class}=87.1
\%,\nonumber\\
P_{{\rm right},2}^{\rm class}=32.4\%,&\qquad&P_{{\rm right},3}^{\rm class}=72.3
\%.
\label{e7}
\end{eqnarray}
The agreement between (\ref{e4}) and (\ref{e7}) is better than that for the
quartic case, possibly because the mixture of successful and failed tunneling
attempts through the central barrier region leads to a more accurate analog of
quantum tunneling. We conclude that complex classical mechanics provides a good
approximation to quantum tunneling for the higher-energy states, as one would
expect of a generalized Bohr correspondence principle.

\begin{figure}
\begin{center}
\includegraphics[scale=0.28, bb=0 0 1000 742]{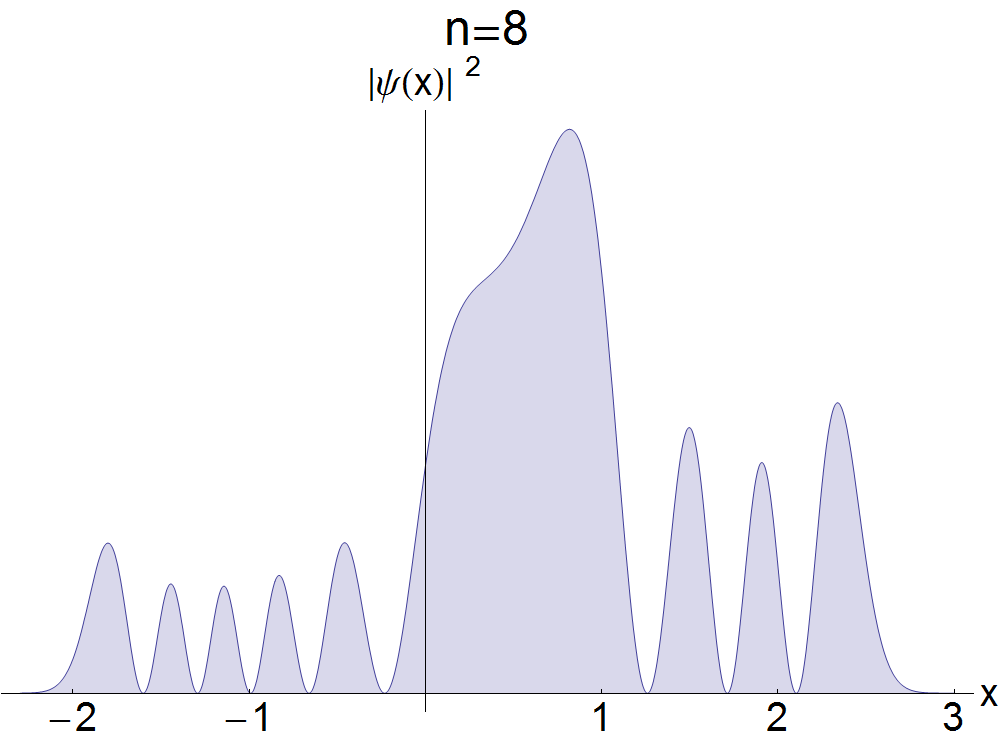}
\end{center}
\caption{Quantum probability density for a particle in the $\psi_8(x)$
eigenstate in the quartic potential (\ref{e1}) with $\hbar=1/2$. The energy of
the particle is $E_8=3.7239$. The probability density has eight nodes. Note that
the particle is most likely to be found inside or near the classically forbidden
region associated with the barrier; this shows that the particle does not have a
high enough quantum number for its behavior to resemble that of a classical
particle.}
\label{F7}
\end{figure}

Of course, the agreements between (\ref{e3}) and (\ref{e6}), and (\ref{e4}) and
(\ref{e7}) are not very precise. To obtain better agreement, it is necessary to
take deeper asymmetric double wells having more quantum eigenenergies below the
top of the central barrier and above the bottom of the shallower well. This can
also be achieved by taking $\hbar$ to have a smaller numerical value. (The
coefficient of $-d^2/dx^2$ in the Schr\"odinger equation is $\hbar^2$.) Until
now, we have taken $\hbar=1$ but we have repeated our numerical work for the
quartic potential (\ref{e1}) with $\hbar=1/2$. Now, the highest eigenvalue below
the top of the barrier is $E_8=3.7239$. The corresponding probability density
$|\psi_8(x)|^2$ is shown in Fig.~\ref{F7}. For this eigenfunction, the quantum
probability of finding the particle to the right of the peak of the barrier is
$60.29\%$ and the corresponding classical probability using the complex energy
$E=E_8+i/4$ is $55.2\%$. (These numbers compare favorably with $P_{{\rm right},
4}^{\rm quant}=40.7689\%$ and $P_{{\rm right},4}^{\rm class}=55.0\%$ for the
highest eigenvalue in the double well with $\hbar=1$.) Doing further
calculations for successively smaller values of $\hbar$ would require a
considerable increase in computer time. The purpose of this paper is only to
demonstrate, in principle, our conjectured connection between complex classical
and quantum tunneling, and not to carry out an extensive and detailed numerical
investigation.

We thank the U.S.~Department of Energy for financial support and Mathematica for
numerical calculations.


\begin{thebibliography}{999}
\bibitem{ZZ} To determine the long-time behavior we calculate numerically the
characteristic time required for a particle to spiral into and away from a
pair of turning points. The characteristic time, which is different for each
well, is almost constant and approaches a constant for very long times.
\end{thebibliography}
\end{document}